\documentclass{appolb}
\usepackage{graphicx}
\usepackage{url}
\usepackage{mathtools}
\usepackage{epsfig,slashed,multirow}
\newcommand{\fsl}[1]{\ensuremath{\mathrlap{\!\not{\phantom{#1}}}#1}}
\def\MG5aMC{{\sc \small MadGraph5\_aMC@NLO}}
\def\FR{{\sc\small FeynRules}}
\def\ML{{\sc\small MadLoop}}
\def\MadFKS{{\sc\small MadFKS}}
\def\MS{{\sc\small MadSpin}}
\def\MW{{\sc\small MadWidth}}
\def\Prospino{{\sc\small Prospino}}
\newcommand{\sss}{\scriptscriptstyle}
\def\Pythia{{\sc\small Pythia}}
\def\MA{{\sc\small MadAnalysis}}

\begin{document}
\title{Colored Particle Production in New Physics at NLO QCD and Its Matching to Parton Showers%
\thanks{Presented at XXIII Cracow EPIPHANY Conference, 9-12 January 2017, IFJ PAN, Cracow, Poland}%
}
\author{Hua-Sheng Shao
\address{Sorbonne Universit\'es, UPMC Univ. Paris 06, UMR 7589, LPTHE, F-75005 Paris, France\\
CNRS, UMR 7589, LPTHE, F-75005 Paris, France}
}
\maketitle
\begin{abstract}
\noindent In this talk, I show the automated Monte Carlo simulations at next-to-leading order in QCD as well as its matching to parton showers are already feasible within the framework of \MG5aMC. I briefly overview the recent activities and take the colored particle production at the LHC as examples. The tools and the models are ready for using by both phenomenologists and experimentalists.
\end{abstract}
\PACS{13.85.-t, 12.38.Bx, 12.60.Jv}
  
\section{Introduction}
The successful operation of LHC provide the unique opportunity to explore the particle physics beyond the Standard Model (BSM) at the TeV scale. Although most of the experimental searches at the LHC do not show any deviations from the Standard Model (SM), it still leaves some possibilities of (in)directly observing new physics at the LHC in the future, either via the precision measurements or via the exotic signatures. The searches at the LHC are heavily relying on the Monte Carlo simulations, where most of the time only leading order (LO) matrix element in a new physics model is taking into account. The main arguments in satisfying the LO results are: 1) there are many possible variations of new physics models; 2) no BSM signatures are confirmed at the LHC; 3) next-to-leading order (NLO) calculations usually take a lot of time on techniques instead of phenomenology. It is true that the first two arguments are still hold today, while the third one start to change with the recent development of automated tools. 

\MG5aMC~\cite{Alwall:2014hca} is such a framework to automated generate NLO accuracy unweighted events after matching to parton showers with the method of {\sc\small MC@NLO}~\cite{Frixione:2002ik}. The infrastructure of \MG5aMC\ requires two pieces of model-dependent counterterms from NLO model generation tools, i.e. the renormalization terms and the rational terms $R_2$~\cite{Ossola:2008xq,Draggiotis:2009yb,Garzelli:2009is,Garzelli:2010qm,Shao:2011tg,Shao:2012ja,Page:2013xla},
where the latter was originally introduced in the OPP method~\cite{Ossola:2006us,Mastrolia:2008jb}. A dedicated module {\sc\small NLOCT}~\cite{Degrande:2014vpa} in \FR~\cite{Christensen:2008py,Alloul:2013bka} was designed to analytically derive the Feynman rules for these counterterms on the fly. Once the NLO model is ready, the virtual contributions (one-loop amplitudes and counterterms) will be taken care of by the module \ML~\cite{Hirschi:2011pa,Ossola:2007ax}, while the real radiation contributions will be computed by \MadFKS~\cite{Frederix:2009yq,Frixione:1995ms}. All of the above procedures are automated as long as the Lagrangian of the model is implemented manually in \FR.

\begin{figure}[htb]
\vspace{-3cm}\centerline{%
\includegraphics[width=12.5cm]{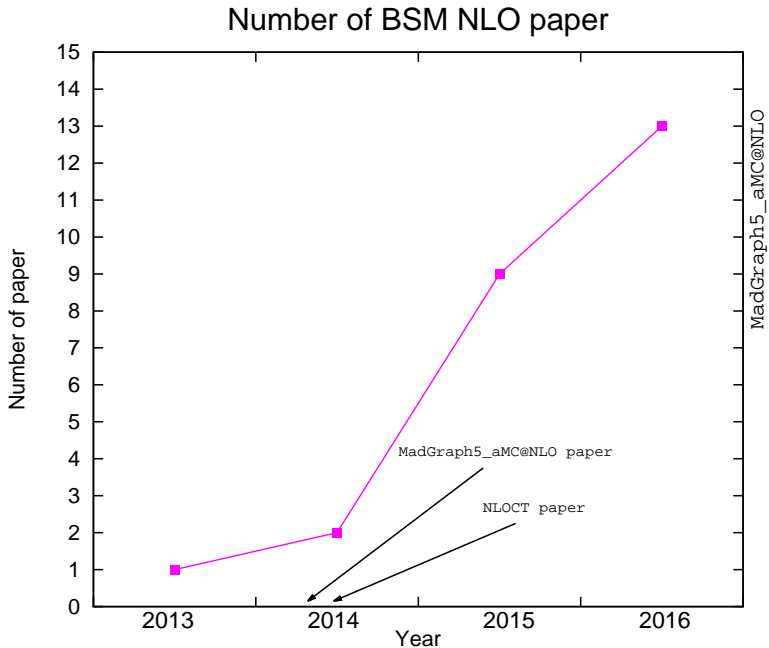}}\vspace{-5cm}
\caption{Summary of number of BSM NLO papers published during 2013-2016 in the \MG5aMC\ framework.}
\label{Fig:BSMNLO}
\end{figure}

Extensive phenomenology studies of NLO QCD corrections in the \MG5aMC\ framework for processes in BSM have been performed, and hence there are already several NLO models available on the \FR\ webpage~\cite{FRWeb}. They are: 
\begin{enumerate}
\item colored particle production in simplified models~\cite{Degrande:2014sta}, supersymmetric QCD~\cite{Degrande:2015vaa} and vector-like quark models~\cite{Fuks:2016ftf}; 
\item Higgs production in the Higgs characterization model~\cite{Artoisenet:2013puc,Maltoni:2013sma,Demartin:2014fia,Demartin:2015uha}, two-Higgs-doublet models~\cite{Degrande:2015vpa,Degrande:2016hyf}, Georgi-Machacek model~\cite{Degrande:2015xnm}; 
\item spin-2 particle production in an effective field theory~\cite{Das:2016pbk};
\item dark matter production at colliders via an s-channel spin-0 or 1 mediator~\cite{Mattelaer:2015haa,Backovic:2015soa,Neubert:2015fka,Arina:2016cqj};
\item top quark production in SM effective field theory up to dimension-6 operators~\cite{Degrande:2014tta,Durieux:2014xla,Bylund:2016phk,Maltoni:2016yxb,Zhang:2016omx,Franzosi:2015osa};
\item Higgs production with dimension-6 operators~\cite{Degrande:2016dqg};
\item heavy neutrino production~\cite{Degrande:2016aje}.
\end{enumerate}
I summarize the number of BSM NLO papers published during 2013-2016 in the \MG5aMC\ framework in Figure~\ref{Fig:BSMNLO}, which clearly shows the increasing activities in the field in recent years.

%uncomment the following lines to place a figure

In this talk, I will only focus on the colored particle production at the LHC. I would like to emphasize that the NLO calculations in BSM are not only trivial applications based on the architectures for the SM calculations. There are many new ingredients should be introduced (e.g. non-renormalized operators, fermion-flow violation, special finite renormalization counterterms etc), which are absent in SM.

\section{Colored particle production}

In this section, I will take three examples of the new colored particle production at the LHC in BSM, which are motivated by the LHC experimental searches. 

\subsection{Colored scalar pair production}

A first example for new colored particle production at NLO QCD accuracy in the \MG5aMC\ framework is the pair production of the color-triplet scalar $\sigma_3$ in a simplified model~\cite{Degrande:2014sta}, which was motived by the stop pair searches at the LHC.~\footnote{In the same paper~\cite{Degrande:2014sta}, we have also studied the color-octet scalar pair production, which was interesting because of the sgluon predictions in the R-parity violation supersymmetric models. I will refrain myself to present these results in this talk.} The corresponding production Lagrangian is
\begin{eqnarray}
\mathcal{L}_3^{\rm p}=D_{\mu}\sigma_3^\dagger D^{\mu}\sigma_3-m_3^2\sigma_3^\dagger\sigma_3,
\end{eqnarray}
while its decay Lagrangian is
\begin{eqnarray}
\mathcal{L}_3^{\rm d}=\frac{i}{2}\bar{\chi}\fsl{\partial}\chi-\frac{1}{2}m_\chi\bar{\chi}\chi+\left[\sigma_3\bar{t}\left(\tilde{g}_LP_L+\tilde{g}_RP_R\right)\chi+{\rm h.c.}\right],
\end{eqnarray}
where $\chi$ is a gauge-singlet Majorana fermion couplings the stop $\sigma_3$ to the top quark. The above simplified model provides the signature of the stop searches in the top quark pair plus missing energy channel. We have analytically checked the expressions of the renormalization counterterms and the $R_2$ terms by three independent calculations. The spin-correlated decays of $\sigma_3$ have been performed with the help of the modules \MS~\cite{Artoisenet:2012st} and \MW~\cite{Alwall:2014bza}. For the illustrative purpose of our numerical analysis, we fixed the new couplings $\tilde{g}_{L,R}$ with the typical values for supersymmetric models featuring a bino-like neutralino and a maximally-mixing top squark
\begin{eqnarray}
\tilde{g}_L=0.25,\tilde{g}_R=0.06.
\end{eqnarray}
The masses of $\sigma_3$ and $\chi$ are kept free. The numerical cross sections have been validated with \Prospino~\cite{Beenakker:1997ut}.

We presented the LO and NLO total cross sections for $pp\rightarrow \sigma_3\bar{\sigma}_3+X$ production at the LHC in Table~\ref{tab:xsec1}. Both 8 TeV and 13 TeV results are given in accompanying with the theoretical uncertainties. In general, the NLO QCD corrections enhance the cross section from $25\%$ to $50\%$ depending on the mass of $\sigma_3$. Scale uncertainties are significantly reduced from $30-40\%$ at LO to $15\%$ at NLO. The PDF uncertainties from 100 NNPDF replicas~\cite{Ball:2012cx} are subdominant except in the large $m_3$ region, because the partonic luminosity is mainly determined by the larger Bjorken fraction $x$ region of the PDF.

\begin{table*}[t]
 \centering
   \begin{tabular}{c||c|c||}
       \multirow{2}{*}{$m_3$ [GeV]} &
       \multicolumn{2}{|c||}{8 TeV} \\
          & $\sigma^{\rm LO}$ [pb] & $\sigma^{\rm NLO}$ [pb]\\
     \hline&&\\[-.4cm] 100 &
          $389.3^{\sss +34.2\%}_{\sss -23.9\%}$ &
          $554.8^{\sss +14.9\%}_{\sss -13.5\%}{}^{\sss +1.6\%}_{\sss -1.6\%}$ \\[.05cm]
      250 &
          $4.118^{\sss +40.4\%}_{\sss -27.2\%}$ &
          $5.503^{\sss +13.1\%}_{\sss -13.7\%}{}^{\sss +3.7\%}_{\sss -3.7\%}$\\[.05cm]
      500 &
          $\big(6.594\times10^{-2}\big){}^{\sss +45.5\%}_{\sss -29.1\%}$ &
          $\big(7.764\times10^{-2}\big){}^{\sss +12.1\%}_{\sss -14.1\%}{}^{\sss+6.7\%}_{\sss -6.7\%}$\\[.05cm]
      750 &
          $\big(3.504\times10^{-3}\big){}^{\sss +48.8\%}_{\sss -30.5\%}$ &
          $\big(3.699\times10^{-3}\big){}^{\sss +12.3\%}_{\sss -14.6\%}{}^{\sss+10.2\%}_{\sss-10.2\%}$ \\[.05cm]
      1000&
          $\big(2.875\times10^{-4}\big){}^{\sss +51.5\%}_{\sss -31.5\%}$ &
          $\big(2.775\times10^{-4}\big){}^{\sss +13.1\%}_{\sss -15.2\%}{}^{\sss +15.5\%}_{\sss -15.5\%}$\\
    \end{tabular}\\[.04cm]
 \begin{tabular}{c||c|c||}
\multirow{2}{*}{$m_3$ [GeV]} &
       \multicolumn{2}{|c||}{13 TeV} \\
          & $\sigma^{\rm LO}$ [pb] & $\sigma^{\rm NLO}$ [pb] \\
     \hline&&\\[-.4cm] 100 &
          $1066^{\sss +29.1\%}_{\sss -21.4\%}$ &
          $1497^{\sss +14.1\%}_{\sss -12.1\%}{}^{\sss +1.2\%}_{\sss -1.2\%}$\\[.05cm]
      250 &
          $15.53^{\sss +35.2\%}_{\sss -24.8\%}$ &
          $21.56^{\sss +12.1\%}_{\sss -12.3\%}{}^{\sss +2.4\%}_{\sss -2.4\%}$ \\[.05cm]
      500 &
          $0.3890^{\sss +39.6\%}_{\sss -26.4\%}$ &
          $0.5062^{\sss +11.2\%}_{\sss -12.8\%}{}^{\sss +4.4\%}_{\sss -4.4\%}$ \\[.05cm]
      750 &
          $\big(3.306\times10^{-2}\big){}^{\sss +41.8\%}_{\sss -27.5\%}$ &
          $\big(4.001\times10^{-2}\big){}^{\sss +10.8\%}_{\sss -12.9\%}{}^{\sss +6.1\%}_{\sss -6.1\%}$ \\[.05cm]
      1000&
          $\big(4.614\times10^{-3}\big){}^{\sss +43.6\%}_{\sss -28.3\%}$ &
          $\big(5.219\times10^{-3}\big){}^{\sss +10.9\%}_{\sss -13.2\%}{}^{\sss +7.9\%}_{\sss -7.9\%}$\\
 \end{tabular}
 \caption{\small \label{tab:xsec1}Total cross sections for $\sigma_3$
   pair production
   at the LHC with the center-of-mass energy $\sqrt{s} = 8$ (upper panel) and $13$~TeV (lower panel). Results are
   presented together with the associated scale and PDF (not shown for
   the LO case) uncertainties. Monte Carlo errors are of about 0.2-0.3\% and omitted. [Table from \cite{Degrande:2014sta}]}
\end{table*}

\begin{figure}[htb]
\centerline{%
\includegraphics[width=12.5cm]{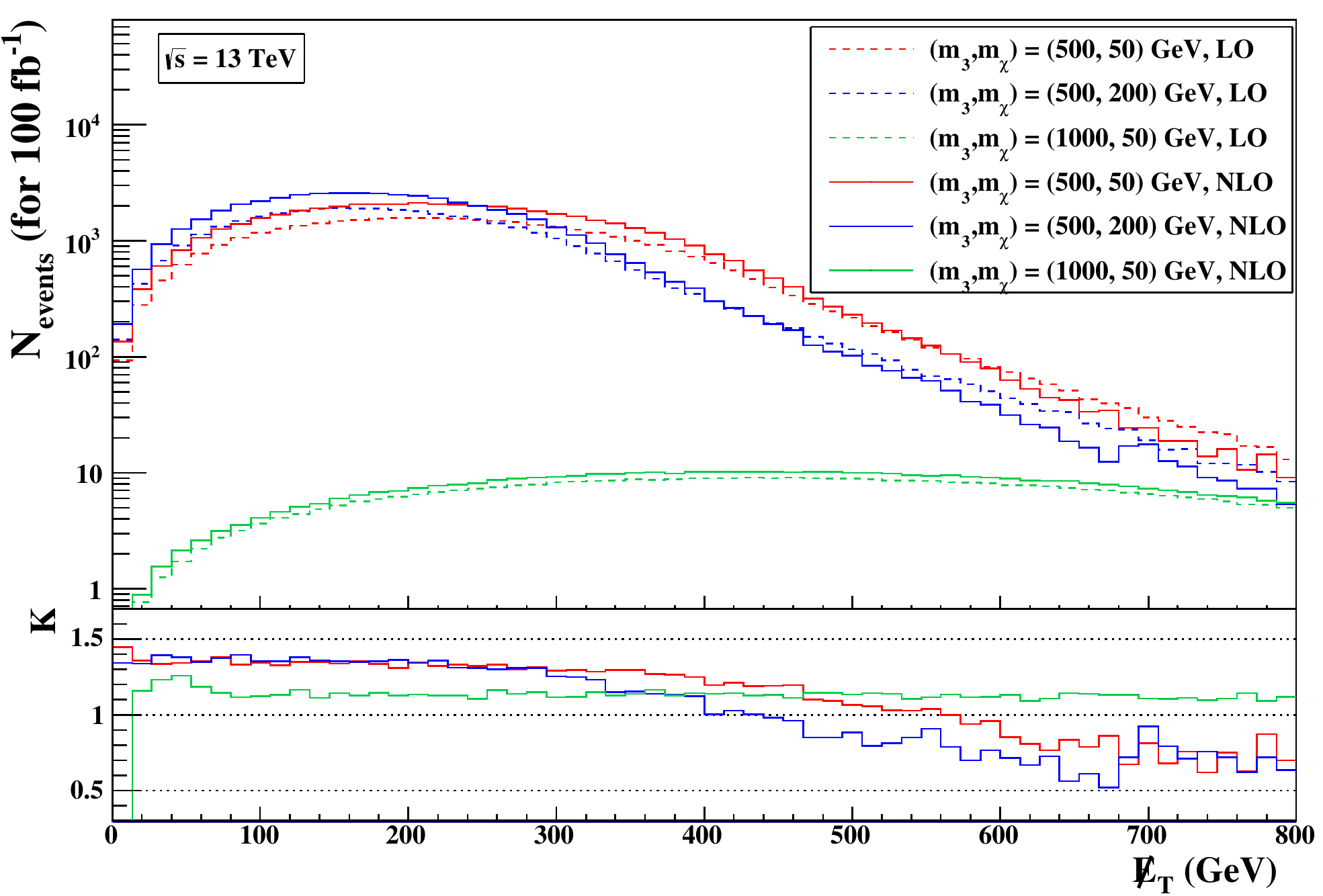}}
\caption{Missing transverse energy distribution for $\sigma_3$ pair production at 13 TeV.[Plot from \cite{Degrande:2014sta}]}
\label{Fig:mets3}
\end{figure}

The inclusive total cross sections are independent of parton showers due to its unitarity. However, the differential distributions may be significantly changed by the parton showers. In the fiducial regions at the LHC, it would be necessary to match NLO QCD calculations with parton showers, which provides exclusive descriptions of the QCD radiations. We examine several differential distributions after matching to parton showers provided by \Pythia8.1~\cite{Sjostrand:2007gs}. The missing transverse energy distributions in three benchmark points $(m_3,m_\chi)=(500,50),(1000,50),(500,200)$ GeV are shown in Figure~\ref{Fig:mets3}, where we imposed single lepton case from the top quark decays in \MA5~\cite{Conte:2012fm}. Although the K-factor depends on the scenario, rescaling a LO sample with a constant K factor are in general not suitable.

\subsection{Supersymmetric QCD}

A second step is to generalize our previous simplified model to the complete supersymmetric QCD~\cite{Degrande:2015vaa}, which includes different flavors of squarks and the gluon's supersymmetric partner gluino. The Lagrangian of the new fields can be described as
\begin{eqnarray}
\mathcal{L}_{\rm SQCD}&=&D_\mu \tilde{q}_L^\dagger D^\mu\tilde{q}_L+D_\mu \tilde{q}_R^\dagger D^\mu\tilde{q}_R+\frac{i}{2}\bar{\tilde{g}}\fsl{D}\tilde{g}-m_{\tilde{q}_L}^2\tilde{q}_L^\dagger\tilde{q}_L-m_{\tilde{q}_R}^2\tilde{q}_R^\dagger\tilde{q}_R-\frac{1}{2}m_{\tilde{g}}\bar{\tilde{g}}\tilde{g}\nonumber\\
&+&\sqrt{2}g_s\left[-\tilde{q}_L^\dagger T \left(\bar{\tilde{g}}P_Lq\right)+ \left(\bar{q}P_L\tilde{g}\right)T\tilde{q}_R+{\rm h.c.}\right]\nonumber\\
&-&\frac{g_s^2}{2}\left[\tilde{q}_R^\dagger T \tilde{q}_R-\tilde{q}_L^\dagger T \tilde{q}_L\right]\left[\tilde{q}_R^\dagger T\tilde{q}_R-\tilde{q}_L^\dagger T\tilde{q}_L\right],
\end{eqnarray}
where the flavor and color indices have been suppressed. Besides the ordinary new renormalization constants, we also need to introduce two new renormalization terms. The first one is the mixing angle renormalization via
\begin{eqnarray}
\left(\begin{array}{l}
\tilde{t}_L\\
\tilde{t}_R
\\
\end{array}\right)
\rightarrow
\left(\begin{array}{l}
\tilde{t}_L\\
\tilde{t}_R
\\
\end{array}\right)+\frac{1}{2}
\left(\begin{array}{ll}
\delta Z_{\tilde{t}_L} & \delta Z_{\tilde{t}_{LR}}\\
\delta Z_{\tilde{t}_{RL}} & \delta Z_{\tilde{t}_{R}}
\\
\end{array}\right)
\left(\begin{array}{l}
\tilde{t}_L\\
\tilde{t}_R
\\
\end{array}\right),
\end{eqnarray}
which generates an off-diagonal mass counterterm
\begin{eqnarray}
\delta \mathcal{L}_{\rm off}=-\delta m^2_{\tilde{t},LR}\left(\tilde{t}^\dagger_L\tilde{t}_R+\tilde{t}^\dagger_R\tilde{t}_L\right).
\end{eqnarray}
Another type of counterterms, the so-called supersymmetric restoring counterterms, should be introduced because we are working in dimensional regularization, which breaks the supersymmetry at loop level. Although there exists dimensional reduction to preserve the supersymmetry beyond tree level, one should still calculate the finite renormalization pieces in order to match with the determination of PDF. We derived these finite supersymmetric restoring counterterms as
\begin{eqnarray}
\mathcal{L}_{\rm SCT}&=&\sqrt{2}g_s\frac{\alpha_s}{3\pi}\left[-\tilde{q}_L^\dagger T \left(\bar{\tilde{g}}P_Lq\right)+\left(\bar{q}P_L\tilde{g}\right)T \tilde{q}_R+{\rm h.c.}\right]\nonumber\\
&+&\frac{g_s^2}{2}\frac{\alpha_s}{4\pi}\left[\tilde{q}_R^\dagger\left\{T_a,T_b\right\}\tilde{q}_R+
\tilde{q}_L^\dagger\left\{T_a,T_b\right\}\tilde{q}_L\right] \left[\tilde{q}_R^\dagger\left\{T^a,T^b\right\}\tilde{q}_R+\tilde{q}_L^\dagger\left\{T^a,T^b\right\}\tilde{q}_L\right]\nonumber\\
&-&\frac{g_s^2}{2}\frac{\alpha_s}{4\pi}\left[\tilde{q}_R^\dagger T_a \tilde{q}_R-\tilde{q}_L^\dagger T_a \tilde{q}_L\right]\left[\tilde{q}_R^\dagger T^a \tilde{q}_R-\tilde{q}_L^\dagger T^a \tilde{q}_L\right].
\end{eqnarray}
Comparing to the full MSSM, we simplified the electroweak sector, while its generalization to the full MSSM is in progress. The decay Lagrangian is
\begin{eqnarray}
\mathcal{L}_{\rm decay}=\frac{i}{2}\bar{\chi}\fsl{\partial}\chi-\frac{1}{2}m_\chi \bar{\chi}\chi+\sqrt{2} g^\prime \left[-\tilde{q}_L^\dagger Y_q \left(\bar{\chi}P_Lq\right)+\left(\bar{q}P_L\chi\right)Y_q\tilde{q}_R+{\rm h.c.}\right].
\end{eqnarray}
In the above equation, $Y_q$ is denoted as the hypercharge of the (s)quarks. 

We studied the case of the gluino pair production in proton-proton collisions, with each gluino decaying into two jets and a $\chi$~\cite{Degrande:2015vaa}. It corresponds to the signature of the multijet production and a missing transverse  energy $\fsl{E}_T$. For simplification, we only consider the scenario of the split supersymmetry~\cite{ArkaniHamed:2004yi} in our numerical analysis, which follows the strategies in ATLAS~\cite{Aad:2014wea,Aad:2015iea} and CMS~\cite{Chatrchyan:2014lfa,Khachatryan:2015vra} gluino searches. The numerical results have been cross checked with \Prospino~\cite{Beenakker:1997ut} in the degenerate squark spectrum case, while our approach is also rigorous in the non-degenerate cases. In Table ~\ref{tab:totalxs2}, we listed the LO and NLO cross sections for the gluino pair production at 13 TeV LHC in terms of the mass of gluino. The NLO QCD corrections enhance the LO cross sections by $50\%$ and reduces the scale uncertainties by a factor of 2. Various differential distributions after matching to parton showers in \Pythia8.2~\cite{Sjostrand:2014zea} via the MC@NLO approach are available in Ref.~\cite{Degrande:2015vaa}.

\begin{table*}[!t]
\renewcommand{\arraystretch}{1.25}
\setlength{\tabcolsep}{12pt}
 \begin{tabular}{c||cc}
    $m_{\tilde{g}}$~[GeV] & $\sigma^{\rm LO}$~[pb] & $\sigma^{\rm NLO}$~[pb] \\
    \hline \hline
   200 & $2104^{+30.3\%}_{-21.9\%}{}^{+14.0\%}_{-14.0\%}$                  &
         $3183^{+10.8\%}_{-11.6\%} {}^{+1.8\%}_{-1.8\%}$ \\
   500 & $15.46^{+34.7\%}_{-24.1\%}{}^{+19.5\%}_{-19.5\%}$                 &
         $24.90^{+12.5\%}_{-13.4\%}{}^{+3.7\%}_{-3.7\%}$ \\
   750 & $1.206^{+35.9\%}_{-24.6\%}{}^{+23.5\%}_{-23.5\%}$                 &
         $2.009^{+13.5\%}_{-14.1\%}{}^{+5.5\%}_{-5.5\%}$\\
  1000 & $1.608\cdot 10^{-1}{}^{+36.3\%}_{-24.8\%}{}^{+26.4\%}_{-26.4\%}$ &
         $2.743\cdot 10^{-1}{}^{+14.4\%}_{-14.8\%}{}^{+7.3\%}_{-7.3\%}$\\
  1500 & $6.264\cdot 10^{-3}{}^{+36.2\%}_{-24.7\%}{}^{+29.4\%}_{-29.4\%}$ &
         $1.056\cdot 10^{-2}{}^{+16.1\%}_{-15.8\%}{}^{+11.3\%}_{-11.3\%}$ \\
  2000 & $4.217\cdot 10^{-4}{}^{+35.6\%}_{-24.5\%}{}^{+29.8\%}_{-29.8\%}$ &
         $6.327\cdot 10^{-4}{}^{+17.7\%}_{-16.6\%}{}^{+17.8\%}_{-17.8\%}$ \\
\end{tabular}
\renewcommand{\arraystretch}{1.0}
\caption{\small \label{tab:totalxs2}LO and NLO QCD inclusive cross sections for
  gluino pair-production at the 13 TeV LHC. The results are shown together with the associated scale
  and PDF relative uncertainties. [Table from \cite{Degrande:2015vaa}]}
\end{table*}

Our sucessful phenomenological application for the Majorana fermion gluino $\tilde{g}$ production indicates the correct treatment of the fermion-flow violation in \MG5aMC, which is a new feature compared to SM physics.

\subsection{Vector-like quark models}

My last example is the vector-like quark production in vector-like quark (VLQ) models~\cite{Fuks:2016ftf}. Vector-like quarks, whose left-handed and right-handed components are in the same representations of the SM gauge groups, are common predictions of many new physics models, like extra dimensions and composite models. They play important roles in CMS and ATLAS experimental searches. The model-independent effective VLQ  Lagrangian is
\begin{eqnarray}
{\mathcal L}_{\rm VLQ} &=&
    i \bar Y \slashed{D} Y - m_{\sss Y} \bar Y Y
  +  i \bar B \slashed{D} B - m_{\sss B} \bar B B
 +  i \bar T \slashed{D} T - m_{\sss T} \bar T T
  +  i \bar X \slashed{D} X - m_{\sss X} \bar X X\nonumber\\
 &-& h\ \bigg[
    \bar B \Big(\hat\kappa_{\sss L}^{\sss B} P_L + \hat\kappa_{\sss R}^{\sss B} P_R\Big) q_d
   + {\rm h.c.} \bigg]- h\ \bigg[
   \bar T \Big(\hat\kappa_{\sss L}^{\sss T} P_L + \hat\kappa_{\sss R}^{\sss T} P_R\Big) q_u
   + {\rm h.c.} \bigg] \\
 &+&
  \frac{g}{2 c_{\sss W}}\ \bigg[
    \bar B\slashed{Z} \Big(\tilde\kappa_{\sss L}^{\sss B}P_L +
      \tilde\kappa_{\sss R}^{\sss B} P_R\Big) q_d + {\rm h.c.} \bigg] +
  \frac{g}{2 c_{\sss W}}\ \bigg[
   \bar T\slashed{Z} \Big(\tilde\kappa_{\sss L}^{\sss T}P_L +
      \tilde\kappa_{\sss R}^{\sss T} P_R\Big) q_u + {\rm h.c.} \bigg] \nonumber\\
 &+&
  \frac{\sqrt{2} g}{2}\ \bigg[
    \bar Y  \slashed{\bar W} \Big(\kappa_{\sss L}^{\sss Y} P_L \!+\! \kappa_{\sss R}^{\sss Y}
       P_R\Big) q_d + {\rm h.c.} \bigg]+
  \frac{\sqrt{2} g}{2}\ \bigg[
   \bar B \slashed{\bar W} \Big(\kappa_{\sss L}^{\sss B} P_L \!+\! \kappa_{\sss R}^{\sss B}
       P_R\Big) q_u + {\rm h.c.} \bigg]\nonumber\\
 &+&
  \frac{\sqrt{2} g}{2}\ \bigg[
   \bar T \slashed{W} \Big(\kappa_{\sss L}^{\sss T} P_L \!+\! \kappa_{\sss R}^{\sss T}
       P_R\Big) q_d + {\rm h.c.} \bigg]+
  \frac{\sqrt{2} g}{2}\ \bigg[
   \bar X  \slashed{W} \Big(\kappa_{\sss L}^{\sss X} P_L \!+\! \kappa_{\sss R}^{\sss X}
        P_R\Big) q_u + {\rm h.c.} \bigg],\nonumber
\end{eqnarray}
where the electromagnetic charges of $(T,B,X,Y)$ are $(+\frac{2}{3},-\frac{1}{3},+\frac{5}{3},-\frac{4}{3})$. The new couplings $\hat{\kappa},\tilde{\kappa},\kappa$ are not fully independent~\cite{Buchkremer:2013bha}
\begin{eqnarray}
\big(\hat\kappa^{\sss Q}_{\sss L,R}\big)_f =&\
     \frac{\kappa_{\sss Q}m_{\sss Q}}{v}
    \sqrt{\frac{\zeta_{\sss L,R}^f\ \xi^{\sss Q}_{\sss H}}
     {\Gamma^{\sss Q}_{\sss H}}} \ , \nonumber\\
 \big(\tilde\kappa^{\sss Q}_{\sss L,R}\big)_f =&\ \kappa_{\sss Q}
    \sqrt{\frac{\zeta_{\sss L,R}^f\ \xi^{\sss Q}_{\sss Z}}
     {\Gamma^{\sss Q}_{\sss Z}}}\ ,\nonumber\\
 \big(\kappa^{\sss Q}_{\sss L,R}\big)_f =&\ \kappa_{\sss Q}
    \sqrt{\frac{\zeta_{\sss L,R}^f\ \xi^{\sss Q}_{\sss W}}
     {\Gamma^{\sss Q}_{\sss W}}}.
\end{eqnarray}
The phenomenological upper limit of the coupling strength $\kappa_{\sss Q}$ can be  loosn to $(0.07,0.2,0.1)$ if the mixing only involves the first, second and third generation respectively.

\begin{figure*}
\centering
 \includegraphics[width=0.80\textwidth]{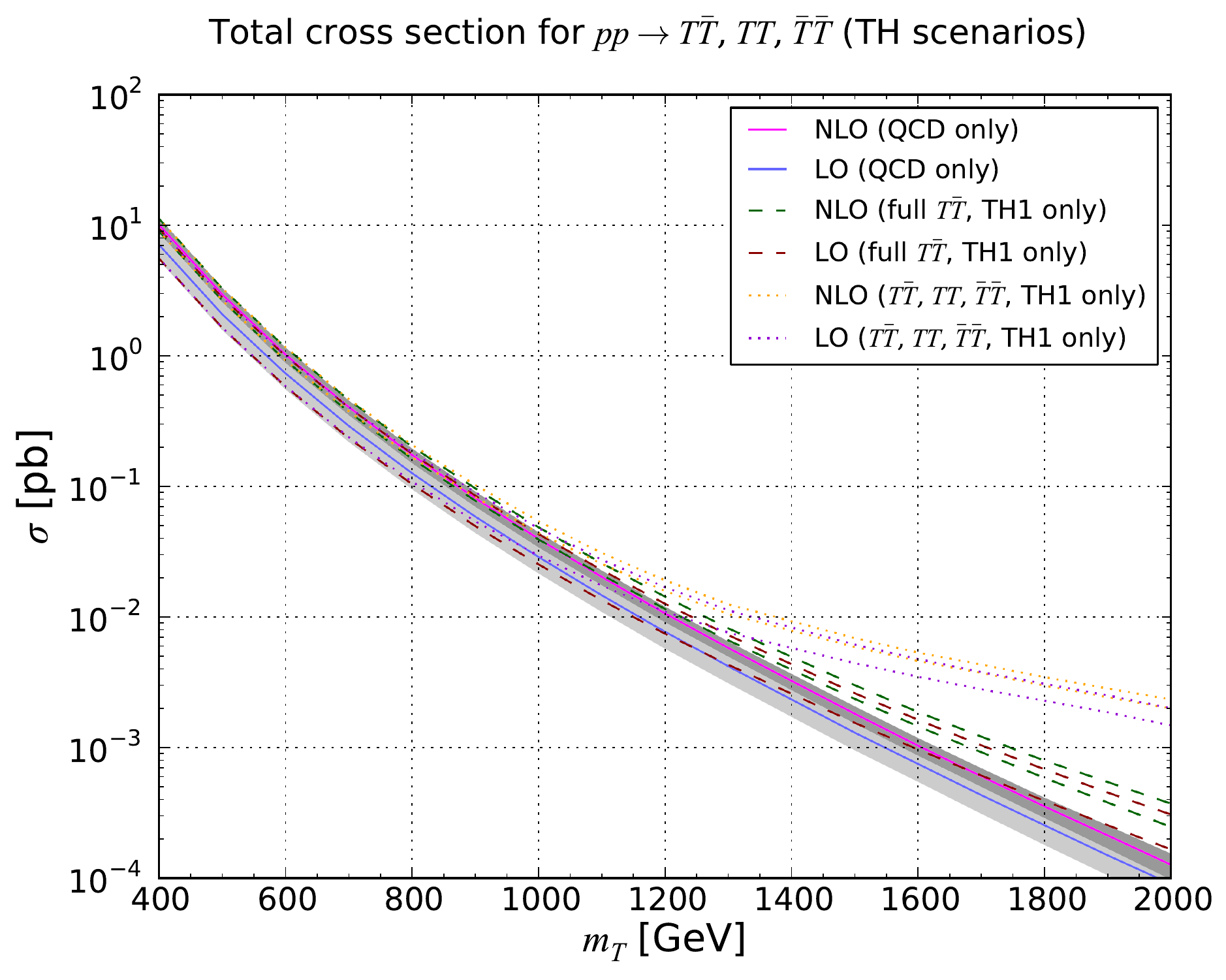}
 \caption{LO and NLO QCD inclusive cross sections for
   $T\bar T$/$TT$/$\bar T \bar T$ pair production at the 13 TeV LHC. The QCD
   contribution results are presented together with the associated theoretical
   uncertainty bands, and we indicate the shifts in the bands that are
   induced by including weak or Higgs-boson exchange diagram contributions when
   they are non-negligible. [Plot from \cite{Fuks:2016ftf}]}
\label{fig:xsec_Tpair}
\end{figure*}

I take VLQ quark $T$ pair  production as an example. The total inclusive cross sections by scanning the mass $m_T$ from 400 GeV to 2 TeV are shown in Figure~\ref{fig:xsec_Tpair}. Due to the kinematical enhancement, the t-channel weak boson exchange diagrams will dominant when $m_T>1.5$ TeV. The same-sign VLQ pair production does not receive any QCD-type Born diagram contributions. Hence, the contribution is only relevant when $m_T>1$ TeV. Extensive discussions and the numerical results can be found in Ref.~\cite{Fuks:2016ftf}.

%uncomment the following lines to place a figure
%\begin{figure}[htb]
%\centerline{%
%\includegraphics[width=12.5cm]{Fig1}}
%\caption{Plot of ...}
%\label{Fig:F2H}
%\end{figure}

\section{Conclusions}

The automation of NLO QCD accuracy simulations in new physics is now feasible via a joint use of \FR\ and \MG5aMC. Many phenomenological studies have been performed in the framework, which help to lift several restrictions applicable only to SM physics. A few NLO models are already available and ready to use for both experimentalists and phenomenologists. I take three examples of new colored particle production at the LHC. In general, NLO QCD corrections are quite sizable and able to reduce the theoretical uncertainties significantly. These results may impact the (re)interpretations of the LHC data in the corresponding new physics models.

\vfill

\noindent{\Large \bf Acknowledgements} \\[24pt]
I acknowledge the supports from ERC grant 291377 ``LHCtheory: Theoretical predictions and analyses of LHC physics: advancing the precision frontier" and the ILP Labex (ANR-11-IDEX-0004-02, ANR-10-LABX-63).

\newpage

\bibliographystyle{utphys}
\bibliography{ref_data}

\providecommand{\href}[2]{#2}\begingroup\raggedright\begin{thebibliography}{10}

\bibitem{Alwall:2014hca}
J.~Alwall, R.~Frederix, S.~Frixione, V.~Hirschi, F.~Maltoni, {\em et al.} {\em
  JHEP} {\bf 1407} (2014) 079,
\href{http://arXiv.org/pdf/1405.0301} 1405.0301.
%%CITATION = ARXIV:1405.0301;%%.

\bibitem{Frixione:2002ik}
S.~Frixione and B.~R. Webber {\em JHEP} {\bf 0206} (2002) 029,
\href{http://arXiv.org/pdf/hep-ph/0204244} hep-ph/0204244.
%%CITATION = HEP-PH/0204244;%%.

\bibitem{Ossola:2008xq}
G.~Ossola, C.~G. Papadopoulos, and R.~Pittau {\em JHEP} {\bf 05} (2008) 004,
\href{http://arXiv.org/pdf/0802.1876} 0802.1876.
%%CITATION = 0802.1876;%%.

\bibitem{Draggiotis:2009yb}
P.~Draggiotis, M.~V. Garzelli, C.~G. Papadopoulos, and R.~Pittau {\em JHEP}
  {\bf 04} (2009) 072,
\href{http://arXiv.org/pdf/0903.0356} 0903.0356.
%%CITATION = 0903.0356;%%.

\bibitem{Garzelli:2009is}
M.~V. Garzelli, I.~Malamos, and R.~Pittau {\em JHEP} {\bf 01} (2010) 040,
\href{http://arXiv.org/pdf/0910.3130} 0910.3130.
%%CITATION = 0910.3130;%%.

\bibitem{Garzelli:2010qm}
M.~V. Garzelli, I.~Malamos, and R.~Pittau {\em JHEP} {\bf 01} (2011) 029,
\href{http://arXiv.org/pdf/1009.4302} 1009.4302.
%%CITATION = 1009.4302;%%.

\bibitem{Shao:2011tg}
H.-S. Shao, Y.-J. Zhang, and K.-T. Chao {\em JHEP} {\bf 09} (2011) 048,
\href{http://arXiv.org/pdf/1106.5030} 1106.5030.
%%CITATION = 1106.5030;%%.

\bibitem{Shao:2012ja}
H.-S. Shao and Y.-J. Zhang {\em JHEP} {\bf 1206} (2012) 112,
\href{http://arXiv.org/pdf/1205.1273} 1205.1273.
%%CITATION = ARXIV:1205.1273;%%.

\bibitem{Page:2013xla}
B.~Page and R.~Pittau {\em JHEP} {\bf 1309} (2013) 078,
\href{http://arXiv.org/pdf/1307.6142} 1307.6142.
%%CITATION = ARXIV:1307.6142;%%.

\bibitem{Ossola:2006us}
G.~Ossola, C.~G. Papadopoulos, and R.~Pittau {\em Nucl. Phys.} {\bf B763}
  (2007) 147--169,
\href{http://arXiv.org/pdf/hep-ph/0609007} hep-ph/0609007.
%%CITATION = HEP-PH/0609007;%%.

\bibitem{Mastrolia:2008jb}
P.~Mastrolia, G.~Ossola, C.~G. Papadopoulos, and R.~Pittau {\em JHEP} {\bf 06}
  (2008) 030,
\href{http://arXiv.org/pdf/0803.3964} 0803.3964.
%%CITATION = 0803.3964;%%.

\bibitem{Degrande:2014vpa}
C.~Degrande {\em Comput. Phys. Commun.} {\bf 197} (2015) 239--262,
\href{http://arXiv.org/pdf/1406.3030} 1406.3030.
%%CITATION = ARXIV:1406.3030;%%.

\bibitem{Christensen:2008py}
N.~D. Christensen and C.~Duhr {\em Comput. Phys. Commun.} {\bf 180} (2009)
  1614--1641,
\href{http://arXiv.org/pdf/0806.4194} 0806.4194.
%%CITATION = 0806.4194;%%.

\bibitem{Alloul:2013bka}
A.~Alloul, N.~D. Christensen, C.~Degrande, C.~Duhr, and B.~Fuks {\em Comput.
  Phys. Commun.} {\bf 185} (2014) 2250--2300,
\href{http://arXiv.org/pdf/1310.1921} 1310.1921.
%%CITATION = ARXIV:1310.1921;%%.

\bibitem{Hirschi:2011pa}
V.~Hirschi, R.~Frederix, S.~Frixione, M.~V. Garzelli, F.~Maltoni, {\em et al.}
  {\em JHEP} {\bf 1105} (2011) 044,
\href{http://arXiv.org/pdf/1103.0621} 1103.0621.
%%CITATION = ARXIV:1103.0621;%%.

\bibitem{Ossola:2007ax}
G.~Ossola, C.~G. Papadopoulos, and R.~Pittau {\em JHEP} {\bf 03} (2008) 042,
\href{http://arXiv.org/pdf/0711.3596} 0711.3596.
%%CITATION = 0711.3596;%%.

\bibitem{Frederix:2009yq}
R.~Frederix, S.~Frixione, F.~Maltoni, and T.~Stelzer {\em JHEP} {\bf 0910}
  (2009) 003,
\href{http://arXiv.org/pdf/0908.4272} 0908.4272.
%%CITATION = ARXIV:0908.4272;%%.

\bibitem{Frixione:1995ms}
S.~Frixione, Z.~Kunszt, and A.~Signer {\em Nucl.Phys.} {\bf B467} (1996)
  399--442,
\href{http://arXiv.org/pdf/hep-ph/9512328} hep-ph/9512328.
%%CITATION = HEP-PH/9512328;%%.

\bibitem{FRWeb}


\bibitem{Degrande:2014sta}
C.~Degrande, B.~Fuks, V.~Hirschi, J.~Proudom, and H.-S. Shao {\em Phys. Rev.}
  {\bf D91} (2015), no.~9, 094005,
\href{http://arXiv.org/pdf/1412.5589} 1412.5589.
%%CITATION = ARXIV:1412.5589;%%.

\bibitem{Degrande:2015vaa}
C.~Degrande, B.~Fuks, V.~Hirschi, J.~Proudom, and H.-S. Shao {\em Phys. Lett.}
  {\bf B755} (2016) 82--87,
\href{http://arXiv.org/pdf/1510.00391} 1510.00391.
%%CITATION = ARXIV:1510.00391;%%.

\bibitem{Fuks:2016ftf}
B.~Fuks and H.-S. Shao {\em Eur. Phys. J.} {\bf C77} (2017), no.~2, 135,
\href{http://arXiv.org/pdf/1610.04622} 1610.04622.
%%CITATION = ARXIV:1610.04622;%%.

\bibitem{Artoisenet:2013puc}
P.~Artoisenet {\em et al.} {\em JHEP} {\bf 11} (2013) 043,
\href{http://arXiv.org/pdf/1306.6464} 1306.6464.
%%CITATION = ARXIV:1306.6464;%%.

\bibitem{Maltoni:2013sma}
F.~Maltoni, K.~Mawatari, and M.~Zaro {\em Eur. Phys. J.} {\bf C74} (2014),
  no.~1, 2710,
\href{http://arXiv.org/pdf/1311.1829} 1311.1829.
%%CITATION = ARXIV:1311.1829;%%.

\bibitem{Demartin:2014fia}
F.~Demartin, F.~Maltoni, K.~Mawatari, B.~Page, and M.~Zaro {\em Eur. Phys. J.}
  {\bf C74} (2014), no.~9, 3065,
\href{http://arXiv.org/pdf/1407.5089} 1407.5089.
%%CITATION = ARXIV:1407.5089;%%.

\bibitem{Demartin:2015uha}
F.~Demartin, F.~Maltoni, K.~Mawatari, and M.~Zaro {\em Eur. Phys. J.} {\bf C75}
  (2015), no.~6, 267,
\href{http://arXiv.org/pdf/1504.00611} 1504.00611.
%%CITATION = ARXIV:1504.00611;%%.

\bibitem{Degrande:2015vpa}
C.~Degrande, M.~Ubiali, M.~Wiesemann, and M.~Zaro {\em JHEP} {\bf 10} (2015)
  145,
\href{http://arXiv.org/pdf/1507.02549} 1507.02549.
%%CITATION = ARXIV:1507.02549;%%.

\bibitem{Degrande:2016hyf}
C.~Degrande, R.~Frederix, V.~Hirschi, M.~Ubiali, M.~Wiesemann, and M.~Zaro
\href{http://arXiv.org/pdf/1607.05291} 1607.05291.
%%CITATION = ARXIV:1607.05291;%%.

\bibitem{Degrande:2015xnm}
C.~Degrande, K.~Hartling, H.~E. Logan, A.~D. Peterson, and M.~Zaro {\em Phys.
  Rev.} {\bf D93} (2016), no.~3, 035004,
\href{http://arXiv.org/pdf/1512.01243} 1512.01243.
%%CITATION = ARXIV:1512.01243;%%.

\bibitem{Das:2016pbk}
G.~Das, C.~Degrande, V.~Hirschi, F.~Maltoni, and H.-S. Shao
\href{http://arXiv.org/pdf/1605.09359} 1605.09359.
%%CITATION = ARXIV:1605.09359;%%.

\bibitem{Mattelaer:2015haa}
O.~Mattelaer and E.~Vryonidou {\em Eur. Phys. J.} {\bf C75} (2015), no.~9, 436,
\href{http://arXiv.org/pdf/1508.00564} 1508.00564.
%%CITATION = ARXIV:1508.00564;%%.

\bibitem{Backovic:2015soa}
M.~Backovi?, M.~KrŠmer, F.~Maltoni, A.~Martini, K.~Mawatari, and M.~Pellen {\em
  Eur. Phys. J.} {\bf C75} (2015), no.~10, 482,
\href{http://arXiv.org/pdf/1508.05327} 1508.05327.
%%CITATION = ARXIV:1508.05327;%%.

\bibitem{Neubert:2015fka}
M.~Neubert, J.~Wang, and C.~Zhang {\em JHEP} {\bf 02} (2016) 082,
\href{http://arXiv.org/pdf/1509.05785} 1509.05785.
%%CITATION = ARXIV:1509.05785;%%.

\bibitem{Arina:2016cqj}
C.~Arina {\em et al.} {\em JHEP} {\bf 11} (2016) 111,
\href{http://arXiv.org/pdf/1605.09242} 1605.09242.
%%CITATION = ARXIV:1605.09242;%%.

\bibitem{Degrande:2014tta}
C.~Degrande, F.~Maltoni, J.~Wang, and C.~Zhang {\em Phys. Rev.} {\bf D91}
  (2015) 034024,
\href{http://arXiv.org/pdf/1412.5594} 1412.5594.
%%CITATION = ARXIV:1412.5594;%%.

\bibitem{Durieux:2014xla}
G.~Durieux, F.~Maltoni, and C.~Zhang {\em Phys. Rev.} {\bf D91} (2015), no.~7,
  074017,
\href{http://arXiv.org/pdf/1412.7166} 1412.7166.
%%CITATION = ARXIV:1412.7166;%%.

\bibitem{Bylund:2016phk}
O.~Bessidskaia~Bylund, F.~Maltoni, I.~Tsinikos, E.~Vryonidou, and C.~Zhang {\em
  JHEP} {\bf 05} (2016) 052,
\href{http://arXiv.org/pdf/1601.08193} 1601.08193.
%%CITATION = ARXIV:1601.08193;%%.

\bibitem{Maltoni:2016yxb}
F.~Maltoni, E.~Vryonidou, and C.~Zhang {\em JHEP} {\bf 10} (2016) 123,
\href{http://arXiv.org/pdf/1607.05330} 1607.05330.
%%CITATION = ARXIV:1607.05330;%%.

\bibitem{Zhang:2016omx}
C.~Zhang {\em Phys. Rev. Lett.} {\bf 116} (2016), no.~16, 162002,
\href{http://arXiv.org/pdf/1601.06163} 1601.06163.
%%CITATION = ARXIV:1601.06163;%%.

\bibitem{Franzosi:2015osa}
D.~Buarque~Franzosi and C.~Zhang {\em Phys. Rev.} {\bf D91} (2015), no.~11,
  114010,
\href{http://arXiv.org/pdf/1503.08841} 1503.08841.
%%CITATION = ARXIV:1503.08841;%%.

\bibitem{Degrande:2016dqg}
C.~Degrande, B.~Fuks, K.~Mawatari, K.~Mimasu, and V.~Sanz
\href{http://arXiv.org/pdf/1609.04833} 1609.04833.
%%CITATION = ARXIV:1609.04833;%%.

\bibitem{Degrande:2016aje}
C.~Degrande, O.~Mattelaer, R.~Ruiz, and J.~Turner {\em Phys. Rev.} {\bf D94}
  (2016), no.~5, 053002,
\href{http://arXiv.org/pdf/1602.06957} 1602.06957.
%%CITATION = ARXIV:1602.06957;%%.

\bibitem{Artoisenet:2012st}
P.~Artoisenet, R.~Frederix, O.~Mattelaer, and R.~Rietkerk {\em JHEP} {\bf 03}
  (2013) 015,
\href{http://arXiv.org/pdf/1212.3460} 1212.3460.
%%CITATION = ARXIV:1212.3460;%%.

\bibitem{Alwall:2014bza}
J.~Alwall, C.~Duhr, B.~Fuks, O.~Mattelaer, D.~G. …ztŸrk, and C.-H. Shen {\em
  Comput. Phys. Commun.} {\bf 197} (2015) 312--323,
\href{http://arXiv.org/pdf/1402.1178} 1402.1178.
%%CITATION = ARXIV:1402.1178;%%.

\bibitem{Beenakker:1997ut}
W.~Beenakker, M.~Kramer, T.~Plehn, M.~Spira, and P.~M. Zerwas {\em Nucl. Phys.}
  {\bf B515} (1998) 3--14,
\href{http://arXiv.org/pdf/hep-ph/9710451} hep-ph/9710451.
%%CITATION = HEP-PH/9710451;%%.

\bibitem{Ball:2012cx}
R.~D. Ball, V.~Bertone, S.~Carrazza, C.~S. Deans, L.~Del~Debbio, {\em et al.}
  {\em Nucl.Phys.} {\bf B867} (2013) 244--289,
\href{http://arXiv.org/pdf/1207.1303} 1207.1303.
%%CITATION = ARXIV:1207.1303;%%.

\bibitem{Sjostrand:2007gs}
T.~Sjostrand, S.~Mrenna, and P.~Z. Skands {\em Comput.Phys.Commun.} {\bf 178}
  (2008) 852--867,
\href{http://arXiv.org/pdf/0710.3820} 0710.3820.
%%CITATION = ARXIV:0710.3820;%%.

\bibitem{Conte:2012fm}
E.~Conte, B.~Fuks, and G.~Serret {\em Comput. Phys. Commun.} {\bf 184} (2013)
  222--256,
\href{http://arXiv.org/pdf/1206.1599} 1206.1599.
%%CITATION = ARXIV:1206.1599;%%.

\bibitem{ArkaniHamed:2004yi}
N.~Arkani-Hamed, S.~Dimopoulos, G.~F. Giudice, and A.~Romanino {\em Nucl.
  Phys.} {\bf B709} (2005) 3--46,
\href{http://arXiv.org/pdf/hep-ph/0409232} hep-ph/0409232.
%%CITATION = HEP-PH/0409232;%%.

\bibitem{Aad:2014wea}
{\bf ATLAS} Collaboration, G.~Aad {\em et al.} {\em JHEP} {\bf 09} (2014) 176,
\href{http://arXiv.org/pdf/1405.7875} 1405.7875.
%%CITATION = ARXIV:1405.7875;%%.

\bibitem{Aad:2015iea}
{\bf ATLAS} Collaboration, G.~Aad {\em et al.} {\em JHEP} {\bf 10} (2015) 054,
\href{http://arXiv.org/pdf/1507.05525} 1507.05525.
%%CITATION = ARXIV:1507.05525;%%.

\bibitem{Chatrchyan:2014lfa}
{\bf CMS} Collaboration, S.~Chatrchyan {\em et al.} {\em JHEP} {\bf 06} (2014)
  055,
\href{http://arXiv.org/pdf/1402.4770} 1402.4770.
%%CITATION = ARXIV:1402.4770;%%.

\bibitem{Khachatryan:2015vra}
{\bf CMS} Collaboration, V.~Khachatryan {\em et al.} {\em JHEP} {\bf 05} (2015)
  078,
\href{http://arXiv.org/pdf/1502.04358} 1502.04358.
%%CITATION = ARXIV:1502.04358;%%.

\bibitem{Sjostrand:2014zea}
T.~Sjšstrand, S.~Ask, J.~R. Christiansen, R.~Corke, N.~Desai, {\em et al.} {\em
  Comput.Phys.Commun.} {\bf 191} (2015) 159--177,
\href{http://arXiv.org/pdf/1410.3012} 1410.3012.
%%CITATION = ARXIV:1410.3012;%%.

\bibitem{Buchkremer:2013bha}
M.~Buchkremer, G.~Cacciapaglia, A.~Deandrea, and L.~Panizzi {\em Nucl. Phys.}
  {\bf B876} (2013) 376--417,
\href{http://arXiv.org/pdf/1305.4172} 1305.4172.
%%CITATION = ARXIV:1305.4172;%%.

\end{thebibliography}\endgroup

\end{document}